\documentclass{aa}  
\usepackage{graphicx}
\usepackage{txfonts}
\begin{document}
   \title{}

   \subtitle{INTEGRAL  and \emph{Swift/XRT} observations of the SFXT  IGR~J16479$-$4514: from quiescence to fast flaring activity} 
   \author{V. Sguera\inst{1},    L. Bassani\inst{1}, R. Landi\inst{1},  A. Bazzano\inst{2}, A. J. Bird\inst{3}, A. J. Dean\inst{3}, 
            A. Malizia\inst{1},  N. Masetti\inst{1}, P. Ubertini\inst{2}
          }
   \offprints{sguera@iasfbo.inaf.it}
   \institute{ IASF/INAF, via Piero Gobetti 101, I-40129 Bologna, Italy  \and 
 IASF/INAF, via Fosso del Cavaliere 100, 00133 Roma, Italy \and 
School of Physics and Astronomy, University of Southampton, Highfield, SO17 1BJ, UK 
         }

   \date{Received 5 December 2007 / Accepted 26 April 2008}


  \abstract
   {IGR~J16479$-$4514 is a fast X-ray transient known to display flares lasting typically a few hours. Recently, its counterpart has been 
    identified with a supergiant star, therefore the source can be classified as member of the newly discovered class of  Supergiant Fast X-ray Transients (SFXTs),
    specifically it is the one with the highest duty cycle.}
   {To characterize the quiescent X-ray behaviour of the source and to  compare  its broad band spectrum to that during fast X-ray flares.} 
   {We performed an analysis of IBIS and JEM-X data with OSA 5.1 as well as an analysis of archival  \emph{Swift}/XRT data.}
    {We present results from a long term monitoring of IGR J16479--4514 with detailed spectral and timing informations on 19 bright fast X-ray flares, 
     10 of which newly discovered. We also report for the first time results on the quiescent X-ray emission;
    the typical luminosity value ($\sim$10$^{34}$ erg s$^{-1}$)  is about 2 orders of magnitude greater than that typical of SFXTs
    while its broad band X-ray spectrum  has a shape very similar to that during fast X-ray transient activity, 
    i.e. a rather steep power law with $\Gamma$$\sim$2.6.}
    {IGR J16479$-$4514 is characterized by a quiescent X-ray luminosity higher than that typical of other known 
    SFXTs but lower than  persistent emission from classical SGXBs.  We suggest that such source is a kind of transition 
    object between these two systems, supporting the idea that there is a continuum of behaviours between 
    the class of SFXTs and that of classical persistent SGXBs.}


   \maketitle 

\section{Introduction}

Since its launch in 2002, the INTEGRAL satellite (Winkler et al. 2003) has played a key role in discovering many new High Mass X-ray Binaries (HMXBs)
thanks to its large field of view (FOV), continous monitoring of the galactic plane and good sensitivity.
The majority of these systems  turned out to be persistent Supergiant High Mass X-ray Binaries (SGXBs) which escaped previous detection because of 
their very obscured nature (e.g. Walter et al. 2006). The remaining ones, named  
Supergiant Fast X-ray Transients (SFXTs,  Negueruela et al. 2006, Sguera et al. 2005, 2006),  were missed before because 
of their very low level of quiescent X-ray luminosities ($\sim$10$^{32}$--10$^{33}$ erg s$^{-1}$) occasionally interrupted by fast X-ray flares 
lasting typically less than a day and reaching peak luminosities of $\sim$ 10$^{36}$ erg s$^{-1}$.  This peculiar transient behaviour was never 
seen before from classical persistent SGXBs which are characterized by X-ray luminosities 
in the range 10$^{36}$--10$^{38}$ erg s$^{-1}$ with  few of them rarely displaying  flaring activity  on few hours timescale, 
i.e. Vela X-1 (Staubert et al. 2004, Laurent et al. 1995). 
As for the physical reason behind fast X-rays flares from SFXTs, they 
should reflect inhomogeneities in the donor star stellar wind  which could be  characterized by a clumpy nature 
(Negueruela et al. 2008, Walter \& Zurita 2007, Leyder et al. 2007).

IGR~J16479$-$4514 is one of the few SFXTs discovered so far by INTEGRAL. It was firstly  detected 
during observations performed between August 8--10 2003 (Molkov et al. 2003); subsequently Sguera et al. (2005, 2006) 
unveiled its fast X-ray transient nature reporting several fast flares strongly resembling 
those of confirmed SFXTs. Recently,  Chaty et al. (2008) and Rahoui et al. (2008)  reported on optical and near/mid infrared observations 
of the source  which lead to the identification of its counterpart with a supergiant star (O8.5I)
at a distance of $\sim$ 4.9 kpc, hence its classification as a SFXT; 
specifically it is the one with the highest duty cycle  so far observed (Sguera et al. 2005, 2006, Walter \& Zurita 2007).

\begin{table*}[t!]
\begin{center}
\caption {Summary of all IBIS detections of fast hard X-ray flares from IGR~J16479$-$4514. The table lists the date 
of their peak emission, approximative duration  of the entire flaring activity, flux and luminosity at the peak (20--60 keV), spectral parameters 
of the bremsstrahlung and power law spectral fits  with their corresponding $\chi^{2}_{\nu}$ and d.o.f. in parenthesis, and finally  reference 
to the discovery paper of each flare.}
\label{tab:main_outbursts}
\begin{tabular}{cccccccc}
\hline
\hline
N. & date            &  duration          &  peak flux    & peak luminosity$\star$          & KT$_{BR}$&  $\Gamma$  & ref \\
    &  (UTC)          &   (hours)         & (20--60 keV, mCrab)      & (10$^{36}$ erg s$^{-1}$) & (keV)    &            &        \\
\hline
1  & 5 Mar 2003, $\sim$14:00  & $\sim$0.5         & $\sim$560                       & $\sim$ 19     & 21$^{+3}_{-2.5}$ (0.56,14) & 2.9$\pm$0.2 (0.87,14)     & 1   \\
2  & 28 Mar 2003, $\sim$8:30  & $\sim$1.5         & $\sim$40                        & $\sim$ 1.3  &         &                            &     1   \\
3  & 21 Apr 2003, $\sim$9:00  &$\sim$0.5$\ddagger$& $\sim$100                      &  $\sim$ 3.4        &  21$^{+6}_{-11}$ (0.6,14) & 3$^{+0.5}_{-0.6}$ (0.54,14) & 1   \\
4  & 10 Aug 2003, $\sim$12:00 & $\sim$60          & $\sim$70                        & $\sim$ 2.4   &                     &    & 4     \\  
5  & 14 Aug 2003, $\sim$1:00  & $\sim$2$\ddagger$ & $\sim$40                       &  $\sim$ 1.3 &   &                      & 1         \\
6  & 11 Aug 2004, $\sim$7:00  & $\sim$8           & $\sim$80                       & $\sim$ 2.7    & 30$^{+15}_{-8}$ (0.74,14)  & 2.5$\pm$0.4 (0.72,14)    & 3    \\
7  & 15 Aug 2004, $\sim$17:00 & $\sim$2           & $\sim$55                        & $\sim$ 1.8    & $\sim$ 40 (1.1,14) & 2.3$\pm$0.9(1.06,14)      &  3  \\
8  & 7  Sep 2004, $\sim$2:00  & $\sim$2           & $\sim$80                        & $\sim$ 2.7      & 44$^{+32}_{-14}$ (0.65,14)  & 2.2$\pm$0.3 (0.65,14)      & 2   \\
9  & 16 Sep 2004, $\sim$17:00 & $\sim$2.5         & $\sim$120                       & $\sim$ 4     &  & 2.6$\pm$0.2 (1.06,14)    & 2    \\
10 & 27 Feb 2005, $\sim$14:30 & $\sim$3           & $\sim$75                       & $\sim$ 2.5     &  46$^{+40}_{-17}$ (0.93,20) & 2.2$\pm$0.45 (1.01,20)     & 3    \\
11 & 27 Mar 2005, $\sim$19:00 & $\sim$1           & $\sim$40                        & $\sim$ 1.3    &   &                     &  3    \\
12 & 3  Apr 2005, $\sim$00:00 & $\sim$2          & $\sim$45                        & $\sim$ 1.5    &   &                      &  3    \\
13 & 4  Apr 2005, $\sim$03:00 & $\sim$9$\ddagger$& $\sim$45                        & $\sim$ 1.5    &  29$^{+15}_{-8}$ (1.02,14)  & 2.6$\pm$0.4 (1.01,14)    & 2    \\
14 & 9  Apr 2005, $\sim$12:00 & $\sim$50          & $\sim$60                        & $\sim$ 2      &   &                       & 3    \\
15 & 12 Aug 2005, $\sim$19:00 &$\sim$9$\ddagger$& $\sim$55                        & $\sim$1.8     &  26$^{+21}_{-8.5}$ (0.6,8)  & 2.5$\pm$0.5 (0.6,8)   &  3    \\
16 & 17 Aug 2005, $\sim$21:30 &  $\sim$8          & $\sim$40                        & $\sim$ 1.3    &          &                       &  3    \\
17 & 26 Aug 2005, $\sim$05:00 &  $\sim$6          & $\sim$40                        & $\sim$ 1.3    &          &                        &  3    \\
18 & 30 Aug 2005, $\sim$04:00 & $\sim$0.5         & $\sim$180                       &  $\sim$ 6     &  38$^{+20}_{-11}$ (0.65,19) & 2.3$\pm$0.35 (0.8,19)  &  5   \\
19 & 3  Mar 2006, $\sim$09:30 &  $\sim$3$\ddagger$ & $\sim$60                        & $\sim$ 2     &          &                       &  3    \\
\hline
\end{tabular}
\end{center}
$\ddagger$ = lower limit on the duration,  $\star$  = assuming a distance of $\sim$ 4.9 kpc (Chaty et al. 2008). \\
(1) Sguera et al. 2005; (2) Sguera et al. 2006; (3) this paper; (4) Molkov et al. 2003; (5) Kennea et al. 2005.
\end{table*}

\begin{table*}[t!] 
\begin{center}
\caption {Summary of \emph{Swift}/XRT  observations of IGR~J16479$-$4514. The table lists the exposure time and the date of 
each observation, the average X-ray flux and the corresponding luminosity (1--9 keV), the photon index $\Gamma$ of the absorbed power law best fit 
spectrum with the corresponding $\chi^{2}_{\nu}$ and d.o.f. in parenthesis, and finally the total absorption  $N_{H}$.}
\begin{tabular}{ccccccc}
\hline
\hline
No.    & exposure  & OBS date&average flux              &  average luminosity$\star$              & $\Gamma$       &  $N_{H}$  \\
       &   ks   & (UTC)  &(1--9 keV, erg cm$^{-2}$ s$^{-1}$) &  (erg s$^{-1}$)                         &                &  10$^{22}$ (cm$^{-2}$)  \\
\hline           
OBS1 (decay flare N. 18) & 0.5 & 30 Aug 2005 &  $\sim$ 1.1$\times$10$^{-10}$   &  $\sim$ 3.2$\times$10$^{35}$    &  $\Gamma$=1.1$\pm$0.8 (0.99,20)  & 8.5$^{+6.5}_{-4.5}$  \\
OBS1 (quiescence) &   8 & 30 Aug 2005   &  $\sim$  6$\times$10$^{-12}$             &  $\sim$ 1.7$\times$10$^{34}$  &  $\Gamma$=0.75$\pm$0.6 (0.45,19) & 4.5$^{+2.2}_{-1.3}$  \\
OBS2  &   6.4  & 10 Sep 2005   & $\sim$ 2.6$\times$10$^{-11}$            &  $\sim$ 7.4$\times$10$^{34}$  &  $\Gamma$=1.35$\pm$0.4 (0.9,54) & 9.5$^{+2.2}_{-1.9}$ \\
OBS3 &   4.1  & 14 Sep 2005   &  $\sim$  2$\times$10$^{-12}$            &  $\sim$ 5.7$\times$10$^{33}$  &  $\Gamma$$\sim$0.6 (0.6,17)        &  $\sim$5         \\
OBS4 &   5.4  & 18 Oct 2005   &  $\sim$ 7.5$\times$10$^{-12}$            &  $\sim$ 2.1$\times$10$^{34}$  &  $\Gamma$=1.8$\pm$0.9 (0.55,17)      & 8.4$^{+5}_{-4}$   \\
\hline        
\end{tabular}
\end{center}
$\star$  = assuming a distance of $\sim$ 4.9 kpc (Chaty et al. 2008, Rahoui et al. 2008).  \\
\end{table*}

Here we report on the characteristics of 10 newly discovered fast flares detected by IBIS and provide for the first time 
20--60 keV spectral information for the set of 19 flares detected so far; for one such flare we also
report and discuss  the broad band X-ray spectrum obtained combining  simultaneous \emph{Swift}/XRT, JEM-X and ISGRI data.
Moreover we present, for the first time, broad band spectral data on the likely quiescent X-ray emission of IGR~J16479$-$4514
which is a very rare information on SFXTs on account of their very recent discovery as class of sources.

\section{Data analysis} 
For this study, we use data collected with IBIS/ISGRI (Ubertini et al. 2003, Lebrun et al. 2003) and JEM-X (Lund et al. 2003), the gamma-ray imager and 
X-ray monitor onboard the INTEGRAL satellite. In particular, the IBIS  data set consists of $\sim$ 2250 pointings or Science Windows 
(ScWs, $\sim$ 2000 seconds duration) where IGR J16479$-$4514 was within 12$^\circ$ from the centre of the instrument FOV. 
All observations were  performed  from approximately the end of 
February 2003 to the beginning of April 2006. ISGRI  
images for each pointing were generated in the 20--60 keV band  using the ISDC offline scientific analysis software OSA
version 5.1; count rates at the position of the source
were extracted from individual images in order to provide the source light curve from which a total 
of 19 flares were identified using the criterium described in section 3. Then an ISGRI spectrum (20--60 keV) 
of each flare was extracted and analysed;  in one case 
(flare N. 18 in Table 1) the source was inside the JEM-X FOV, so that low energy spectral data could 
also be obtained over the 4--20 keV band.
In order to further study the soft X-ray properties of the source, we also used  X-ray data collected 
with XRT (X-ray Telescope) on board the \emph{Swift}  satellite (Gehrels et al. 2004) whenever available. 
From  the \emph{Swift} archive, we found that IGR J16479$-$4514 
was observed four times during the period August-October 2005; Table 2 reports for each observation the corresponding exposure, date,
X-ray flux and luminosity  (1--9 keV) as estimated using an absorbed power law fit to the XRT data. 
XRT data reduction was performed  according to the processes described in Landi et al. (2007).
All spectral analysis reported in the paper was performed  using \emph{Xspec}  version 11.3; uncertainties are given at the 90\%
confidence level for one single parameter of interest.

\begin{figure}[t!]
\centering
\includegraphics[width=7.2cm,height=9.5cm,angle=270]{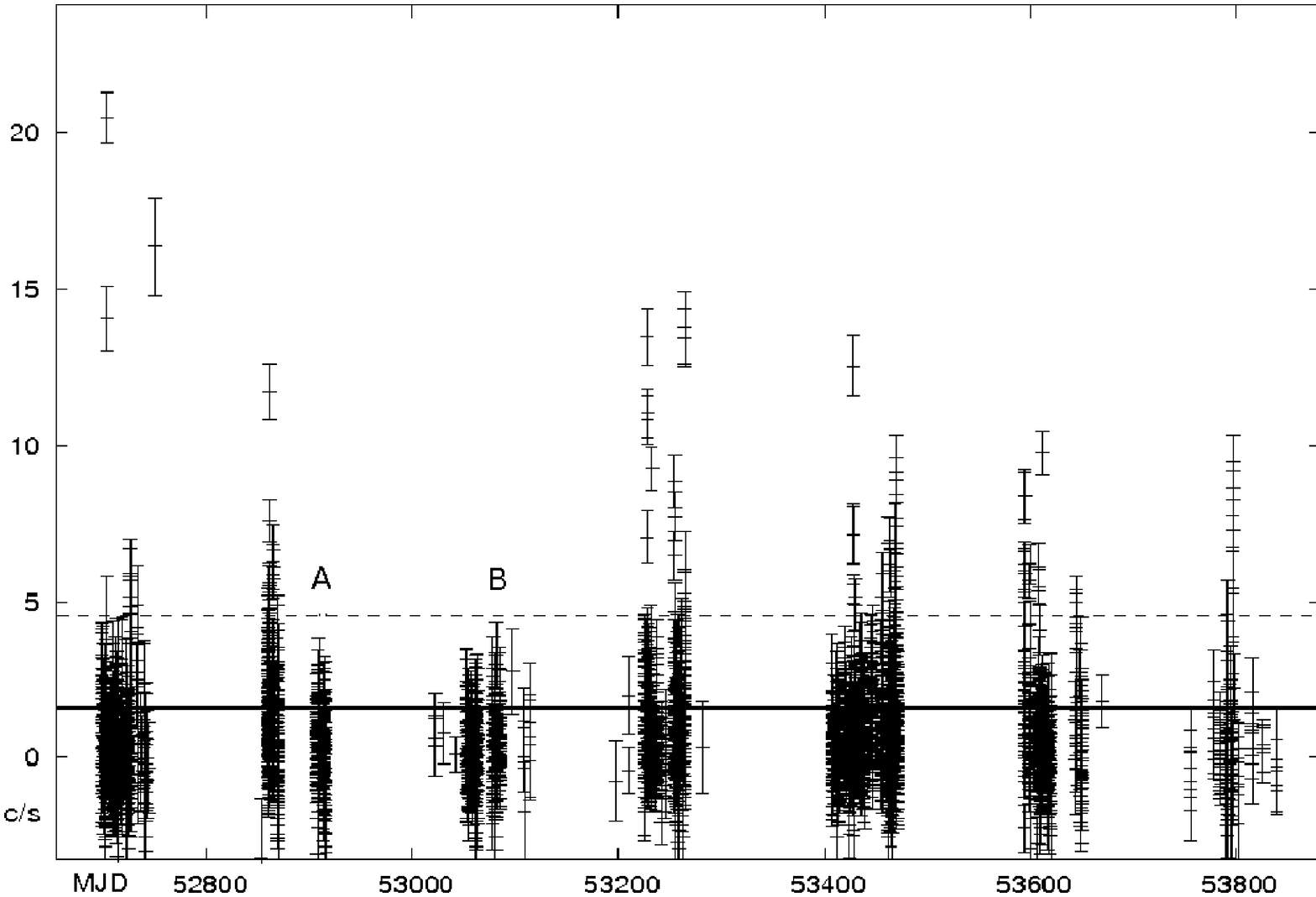}
\caption{ISGRI long term light curve (20--60 keV) of IGR~J16479$-$4514. Time and flux axis are in MJD and count s$^{-1}$, respectively.
Each data point represents the average flux during one ScW ($\sim$ 2000 seconds).}
\centering
\includegraphics[width=9.1cm,height=6.8cm]{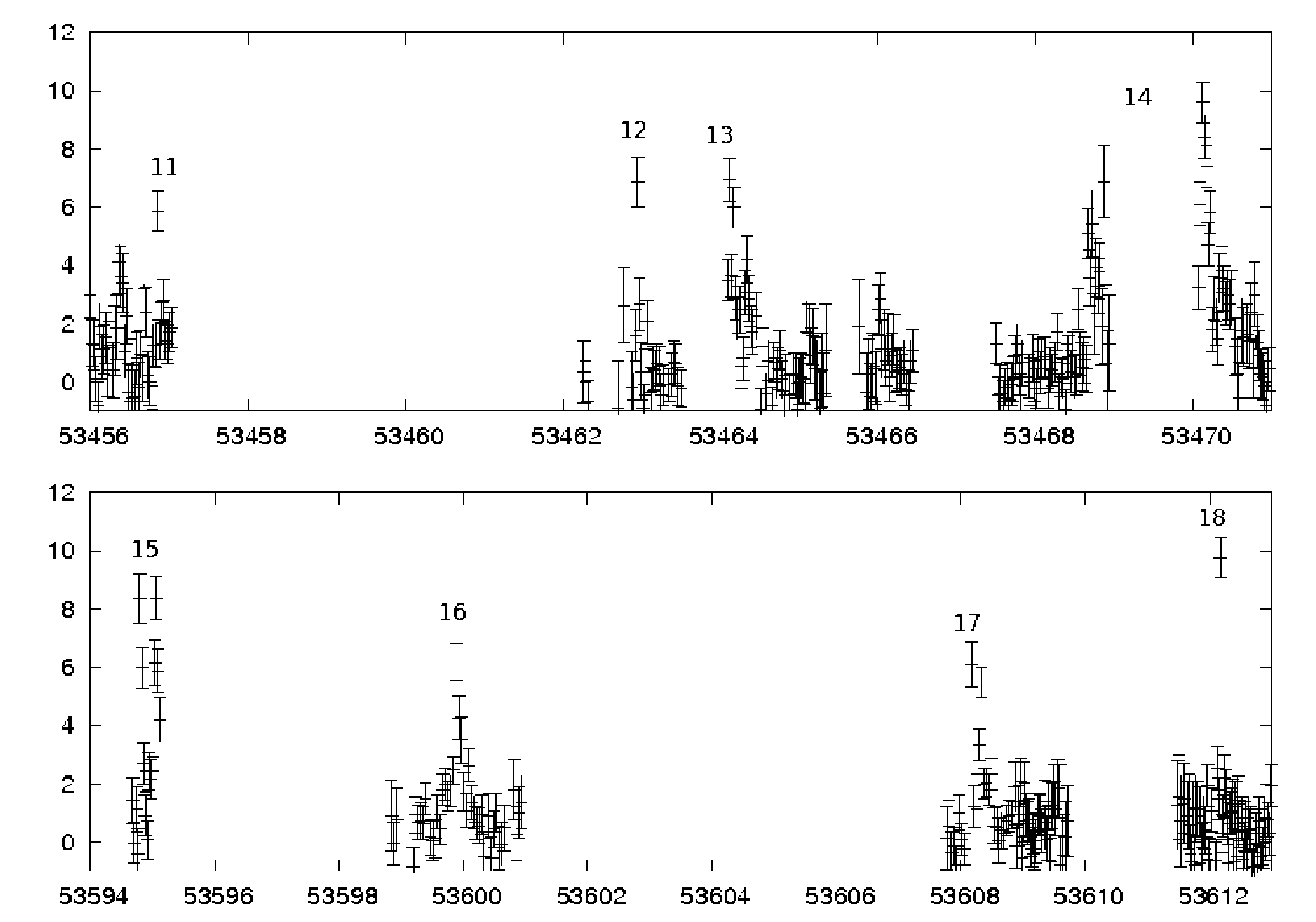}
\caption{Zoomed views of the light curves of flares from N. 11 to 18. Time and flux axis are in MJD and count s$^{-1}$, respectively.
Each data point represents the average flux during one ScW ($\sim$ 2000 seconds).}
\end{figure}

\section{Timing analysis}

The ISGRI long term light curve (20--60 keV) of  IGR J16479$-$4514 on ScW timescale is shown in Fig. 1, where 
the black line  represents the 2$\sigma$ upper limit at the ScW level ($\sim$10 mCrab or 1.2$\times$ 10$^{-10}$ erg cm$^{-2}$ s$^{-1}$).
Most of the time the source is not significantly detected at ScW level and it is below the instrumental sensitivity of ISGRI; sporadically it undergoes 
fast X-ray flares. In particular we considered  those outbursts having a peak flux greater than $\sim$ 30 mCrab or 
3.6$\times$10$^{-10}$ erg cm$^{-2}$ s$^{-1}$ (20--60 keV); this peak flux value is represented in Fig. 1 by  the broken line and  
corresponds to a source significance detection equal to or greater than $\sim$ 6$\sigma$ in the single ScW containing the peak of the flare.
By applying this criterium, a total of 19 fast X-ray flares have been detected and they are listed in Table 1 together 
with the date of the peak emission, approximative duration  of the entire flaring activity, flux and luminosity at 
the peak (20--60 keV). In particular, ten new flares are reported here for the first time (N. 6,7,10,11,12,14,15,16,17,19 in Table 1). 
We note that Walter \& Zurita (2007) 
have reported a total of 38 flares (27 short and 11 long) from this source although no detailed analysis of individual flares is presented; 
the difference in the number of flares is likely due to a different total observing time and flare definition. In particular, we adopt a conservative peak flux threshold 
for flare recognition in order to pick up flare bright enough to extract a meaningful ISGRI  spectrum.
 
On the basis of Table 1, although the typical flare duration is only a few hours, we note that 
the source occasionally displays activity over a period of a few days. Following the classification  between short and long flares by  Walter \& Zurita (2007),
we report 12 flares of the first type and 7 of the second, i.e a similar ratio as they reported.  
The typical peak flux is in a narrow range $\sim$40--80 mCrab (20--60 keV) but occasionally much brighter flares  occur. 
A detailed analysis of Fig. 1 indicates that the typical flare recurrence time is $\sim$1-2 days: 
19 flares were detected over a total exposure of $\sim$ 23 days while occasionally more flares ($\sim$4) were seen over a $\sim$ 9 days period 
(see examples in Fig. 2). This agrees quite well with 
the average recurrence time reported by Walter \&  Zurita (2007).  To search for real evidence of periodicity,
 we further used  the Lomb-Scargle method with the fast implementation 
of Press \& Rybicki (1989) and Scargle (1982) but no indication of  periodicity was found in the range  1--300 days. We also searched a 1 second bin time ISGRI  light curve 
of all brightest outbursts in Table 1 (N. 1,6,9,10,18) for pulsations but none were found.

A deeper inspection of Fig. 1 also shows that  the source occasionally enters long periods of very low flux level (for example, blocks
A and B in Fig. 1) and  we tentatively associate them to the source quiescence which must be below $\sim$ 10 mCrab (20--60 keV).

Flare N. 18 in Table 1 is particularly interesting because it was discovered by 
\emph{Swift}/BAT (15--50 keV)  which promptly triggered the \emph{Swift}/XRT observation (1--8 keV) that caught 
the flare only during its decay phase (Kennea et al. 2005). 
Here we report for the first time on the simultaneous JEM-X and ISGRI detection of such flare which provides a light curve (200 seconds bin time) 
in two different energy bands:  10--20 keV  (JEM-X) and 20--60 keV (ISGRI) as reported in Fig. 3.

During the four \emph{Swift}/XRT observations (see Table 2), IGR J16479$-$4514  showed flaring activity 
only at the beginning of OBS1 (see bottom Fig. 3) where two X-ray flares  are evident. 
Apart from the first flare previously cited, a second one started $\sim$ 4,000 seconds later with a duration of $\sim$ 1,000 seconds.
It was outside the JEM-X FOV while it may be present in the ISGRI light curve (see top Fig. 3) however without sufficient statistical significance for a secure claim
because it was too faint  (average flux $\sim$ 2$\times$10$^{-11}$ erg cm$^{-2}$ s$^{-1}$, 1--9 keV).
No more flaring activity was detected in the remaining part of  OBS1 nor in the following  \emph{Swift}/XRT pointings (OBS2/3/4) and 
the source appeared to have reached its likely quiescent state
with a typical 1--9 keV luminosity  of $\sim$ 10$^{34}$ erg s$^{-1}$ (see Table 2).

\section{X-ray spectral analysis}

\begin{figure}[t!]
\includegraphics[width=9cm,height=6.5cm]{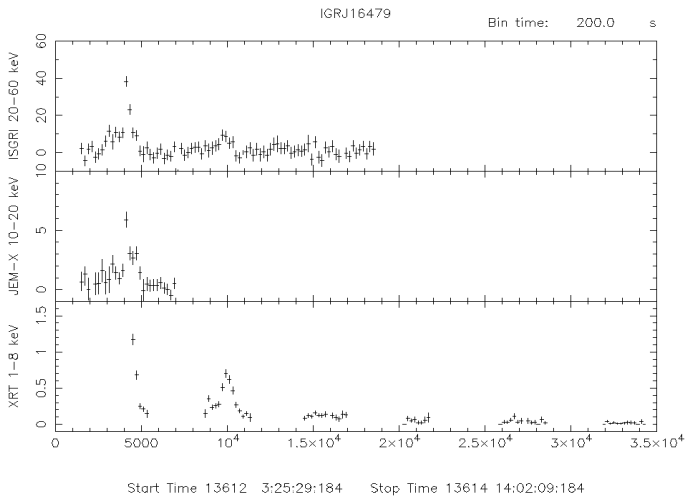}
\caption{ISGRI (top), JEM-X (middle) and \emph{Swift/XRT} (bottom) simultaneous light curves of the flare N. 18 in Table 1. The bin time is 200 seconds.}
\end{figure}

\subsection{Flares}
To date ISGRI spectral information on flares from SFXTs is sparse, however  in the case of  IGR J16479$-$4514 the 
large number of bright flares allows a proper study to be performed. For the majority of flares reported in Table 1 we were able to 
extract an ISGRI spectrum and perform a fit with two different spectral models: power law and bremsstrahlung. Spectral parameters, 
$\chi^{2}_{\nu}$  and corresponding  degree of freedom (d.o.f) are all listed in Table 1.
A discrimination between these two models on a statistical basis is not possible because all fits 
give acceptable and comparable $\chi^{2}_{\nu}$; 
the bremsstrahlung temperatures and power law indices fall in a narrow range of kT=21--46 keV and $\Gamma$=2.2--3 and this
suggests  constancy in shape (but not flux)  of the source spectra from flare to flare. 
Bearing this in mind and  in order to improve the statistics, ISGRI spectra from all flares
were fit together using the two models previously adopted; the bremsstrahlung model provided a kT=27$^{+3.2}_{-3.6}$ keV
($\chi^{2}_{\nu}$=0.9, 109 d.o.f.) while  the power law  gave a  $\Gamma$=2.66$\pm$0.13
($\chi^{2}_{\nu}$=0.95, 109 d.o.f.). 

Next we analysed in detail flare N. 18 using all avaliable data. An absorbed power law fit to the   \emph{Swift}/XRT 
data alone (relative to only the first flare in bottom Fig. 3) provided a flat photon index and absorption N$_{H}$ (see Table 2)
in excess to the galactic value, which along the line of sight is 2.1$\times$10$^{22}$ cm$^{-2}$ (Dickey \& Lockman 1990).
JEM-X data alone are also well fit by  a power law with photon index $\Gamma$=2.2$\pm$0.2, a value 
very similar to that found by ISGRI. 
A change in shape is clearly evident going from \emph{Swift}/XRT to JEM-X/ISGRI, 
possibly due to extra absorption or to a high energy  cutoff. We subsequently performed the broad band spectral analysis over the 2--60 keV energy range;
a simple power law model poorly fit the data since the residuals clearly show the presence of absorption 
at soft X-rays. In fact, an absorbed power law provides a meaningful  fit to the data ($\chi^{2}_{\nu}$=0.9, 154 d.o.f.) 
with $\Gamma$=2.5$\pm$0.2  and N$_{H}$=16$\pm$3$\times$10$^{22}$ cm$^{-2}$, Fig. 4 (top) 
displays such unfolded broad band spectrum.  No cut-off is statistically required in the data fit.
We introduced a costant to take into account possible miscalibrations between the instruments employed 
as well as to compensate for the incomplete coverage of \emph{Swift}/XRT.
The XRT/JEM-X and JEM-X/ISGRI constants were found to be 0.13$^{+0.25}_{-0.02}$ and 1$^{+0.3}_{-0.2}$, respectively;  
the former is low because  \emph{Swift}/XRT detected the flare only during its decay phase while JEM-X/ISGRI detected it  throughout the entire duration of the event.
The  total  N$_{H}$ inferred from the absorbed power law  broad band fit is much higher than that obtained from individual
\emph{Swift}/XRT spectral fits; such high N$_{H}$ could explain why the  \emph{Swift}/XRT spectrum have a rather hard photon index compared to the steeper JEM-X/ISGRI spectra.
Moreover, an absorbed bremsstrahlung model fits equally well  
the data ($\chi^{2}_{\nu}$=1.05, 154 d.o.f., kT=19$^{+6.3}_{-4.3}$,  N$_{H}$=9$^{+2.5}_{-2}$$\times$10$^{22}$ cm$^{-2}$)
and the values of the two constants are very similar to those previously found.

\begin{figure}[t!]
\includegraphics[width=8.6cm,height=5.5cm]{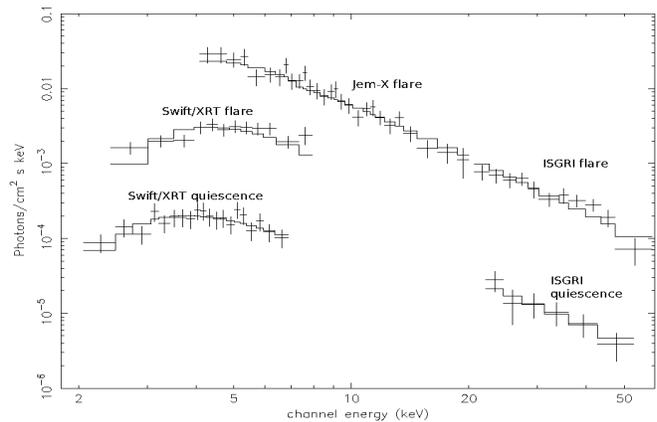}
\caption{Unfolded broad band spectrum (2--60 KeV) of flare N.18 in Table 1 (top) and of the quiescent emissiom (bottom).}
\end{figure}

\subsection{Quiescence}
From the long term light curve of IGR J16479$-$4514, we individuated a total of $\sim$ 530 pointings during which 
the source is not significantly detected in any individual ScW (see blocks A and B in Fig. 1); 
however a mosaic of all these ScWs provided a clear detection of the source  
at $\sim$ 9$\sigma$ level in the 20--60 keV band.
The fluxes for spectral analysis were extracted from the location of IGR J16479$-$4514  in fine band mosaics of all  530 ScWs and a
spectrum was produced according to the processes described in Bird et al. (2007).
This spectrum is equally well fit using a power law ($\Gamma$=2.5$\pm$1, $\chi^{2}_{\nu}$=0.2, 4 d.o.f.) and  
a bremsstrahlung (kT=30$^{+47}_{-14}$ keV, $\chi^{2}_{\nu}$=0.2, 4 d.o.f.).
The average 20--60 keV flux and luminosity are $\sim$ 1.7$\times$10$^{-11}$ erg cm$^{-2}$ s$^{-1}$ 
and   $\sim$5$\times$10$^{34}$ erg s$^{-1}$; Walter \& Zurita (2007) 
reported a similar ISGRI quiescent flux of $\sim$1.54$\times$10$^{-11}$ erg cm$^{-2}$ s$^{-1}$ confirming that such value 
is the lowest hard X-ray emission level detected from the source to date.

The soft X-ray properties of the quiescence as detected by  \emph{Swift}/XRT can be safely associated to OBS2/3/4 and to the final part of OBS1.
All spectra pertaining to these observations are best fit by an absorbed power law model (see Table 2); 
the values of  the photon index are compatible within the uncertainties indicating that the source may be 
characterized by the same rather hard spectral shape during quiescence and also during the decay phase of flare N. 18.
We checked for variability in the spectral indices between OBS4 and  OBS1/quiescence by fixing the $N_{\rm H}$ value of 
the first observation to that of the second; by doing so, no variability 
has been found since the photon index assumes an almost identical value of $\Gamma$=$0.9\pm0.3$.

We combined the \emph{Swift}/XRT spectrum of OBS4 with that of  ISGRI in quiescence  
to obtain broad band energy information  over the 2--60 keV band. The underlyng assumption is that the 
spectral shape of the source did not change during the time interval between the \emph{Swift}/XRT and the  IBIS 
observations, which is reasonable given the constancy in shape seen both by IBIS and \emph{Swift}/XRT.
The best fit is provided by an absorbed power law model ($\chi^{2}_{\nu}$=0.5, 22 d.o.f.) 
with  $\Gamma$=2.2$\pm$0.75 and a  total N$_{H}$=9$^{+4.0}_{-3.5}$$\times$10$^{22}$ cm$^{-2}$; the cross calibration 
constant between the two instruments is 0.5$^{+1}_{-0.3}$.  Fig. 4 (bottom) displays the unfolded broad band spectrum.
Also in this case, the bremsstrahlung provided a comparably good fit ($\chi^{2}_{\nu}$=0.4, 22 d.o.f.), however the temperature
was not well constrained (kT$\sim$ 20 keV). We point out that the broad band X-ray spectral shapes of the source 
in quiescence and during flaring activity  are very similar, as it can be clearly noted in Fig. 4.  

\section {Discussion and conclusions}

In this paper, we present results from a long term monitoring of IGR J16479$-$4514 with detailed informations on 19  bright flares, 10 of which were newly discovered.  
The flares are detected typically every $\sim$1-2 days, this makes IGR J16479$-$4514  the SFXT with the 
highest duty cycle so far seen. However, occasionally the source is in a quiescent state; the longest period of inactivity sampled by our
monitoring is $\sim$ 12 days. The typical flare duration is only a few hours but occasionally longer flaring activity has been 
detected. Our detailed X-ray spectral 
analysis shows that the shape of the source in quiescence and during flares  is identical (i.e a rather steep power law  
with $\Gamma$$\sim$2.6), despite large  
excursions in flux. Moreover, the source is always detected when observed  by an X-ray instrument having sufficient sensitivity, such as \emph{Swift}/XRT, 
with a 1--9 keV flux (luminosity)  of $\sim$ 10$^{-12}$  erg cm$^{-2}$ s$^{-1}$ ($\sim$ 10$^{34}$ erg s$^{-1}$). Since the typical peak flux  of flares is  $\sim$ 10$^{-9}$ 
erg cm$^{-2}$ s$^{-1}$, IGR J16479$-$4514 must accrete over 
a dynamic range of $\sim$ 3,500. More importantly, its quiescence is higher than that typical of other SFXTs
by about two orders of magnitude (Negueruela et al. 2006) and this raises the possibility that during the quiescence 
of  IGR J16479$-$4514, the compact object is still close to the supergiant donor
star and so it is still accreting a significant amount of material from its wind but 
not in the form of clumps as during fast X-ray flares. This would explain the same spectral shape seen 
in quiescence and during flares  since the emission  mechanism should be the same, i.e. accretion onto the compact object.
Consecutively, the system should be characterized by a small  orbital radius and weak eccentricity, 
otherwise it would be in quiescence for longer intervals ($>$ 12 days) and with lower X-ray luminosity than those effectively observed.

Considering that the  quiescent X-ray luminosity of IGR J16479$-$4514 is intermediate between  that of other SFXTs 
and classical persistent SGXBs, we suggest that this system  is 
a transition object between the two classes; this  supports the idea that there is a continuum of behaviours between SFXTs and classical SGXBs (Negueruela et al. 2007). 
What differentiates such systems is most likely the different wind properties and/or orbital parameters,
i.e. orbital radius, orbital period and eccentricity (Negueruela et al. 2008, Chaty et al. 2008). 
It is important to point out that in the literature there are 
at least three other unidentified X-ray sources which could be similar to  IGR J16479$-$4514, 
i.e. IGR J16195$-$4945 ( Sguera et al. 2006, Walter \& Zurita 2007, Tomsick et al. 2006), IGR J16418$-$4532 (Sguera et al. 2006, Walter et al. 2006) and  
XTE J1743$-$363 (Sguera et al. 2006, Walter \& Zurita 2007).

The accumulation of exposure time and longer temporal coverage of the source by IBIS 
is very likely to further increase the possibility of discovering the  orbital period 
of  IGR J16479$-$4514 which 
could provide a key information to study and understand 
the physical reasons behind its very unusual X-ray behaviour.

\begin{acknowledgements}
The authors acknowledge the ASI financial support via grant
ASI-INAF I/088/06/0, ASI-IANF I/023/05/0

\end{acknowledgements}

\end{document}